\documentclass[12pt]{article}
\usepackage{graphicx}

\newcommand\pubnumber{SNSN-323-63}
\newcommand\pubdate{\today}

\def\napoli{$^{*}$Department of Physics, Faculty of Science, Kyoto University,\\
Kitashirakawa, Oiwake-cho, Sakyo-ku, Kyoto-shi, Kyoto 606-8502, Japan}

\def\Title#1{\begin{center} {\Large #1 } \end{center}}
\def\Author#1{\begin{center}{ \sc #1} \end{center}}
\def\Address#1{\begin{center}{ \it #1} \end{center}}

\newcommand\pubblock{\rightline{\begin{tabular}{l} \pubnumber\\
         \pubdate  \end{tabular}}}
\newenvironment{Abstract}{\begin{quotation}  }{\end{quotation}}
\newenvironment{Presented}{\begin{quotation} \begin{center} 
             PRESENTED AT\end{center}\bigskip 
      \begin{center}\begin{large}}{\end{large}\end{center} \end{quotation}}
\def\Acknowledgements{\bigskip  \bigskip \begin{center} \begin{large}
             \bf ACKNOWLEDGEMENTS \end{large}\end{center}}

\def\beq{\begin{equation}}
\def\eeq#1{\label{#1}\end{equation}}
\def\eeqn{\end{equation}}

\def\beqa{\begin{eqnarray}}
\def\eeqa#1{\label{#1}\end{eqnarray}}
\def\eeqan{\end{eqnarray}}

\let\bar=\overbar

\def\Dslash{\not{\hbox{\kern-4pt $D$}}}
\def\dslash{\not{\hbox{\kern-2pt $\del$}}}

\def\msb{{\bar{\ssstyle M \kern -1pt S}}}

\begin{document}
\begin{titlepage}
\pubblock

\vfill
\Title{Improving Charge-Collection Efficiency of Kyoto's \\SOI Pixel Sensors}
\vfill
\Author{Hideaki Matsumura$^{*}$, T. G. Tsuru, T. Tanaka, A. Takeda, M. Ito,\\ 
S. Ohmura, Y. Arai, K. Mori, Y. Nishioka, R. Takenaka, T. Kohmura} 
\Address{\napoli}
\vfill
\begin{Abstract}
We have been developing X-ray SOIPIXs for next-generation satellites for X-ray astronomy.
Their high time resolution ($\sim10~\mu$s) and event-trigger-output function enable us to  read out without  pile-ups and to use  anti-coincidence systems.
 Their performance in imaging spectroscopy is comparable to that in the CCDs.
A problem  in our previous  model was degradation of charge-collection efficiency (CCE) at pixel borders.
We measured the  response in the sub-pixel scale,  using finely collimated X-ray beams at $10~\mu$m$\Phi$ at SPring-8,  and investigated the non-uniformity of the CCE within a pixel.
We found that the X-ray detection efficiency and CCE degrade in the sensor region under the pixel circuitry placed outside the buried p-wells (BPW).
A 2D simulation of the electric fields shows that the isolated pixel-circuitry outside the BPW  creates local minimums in the electric potentials at the interface between the sensor and buried oxide layers.
 Thus, a part of signal charge is trapped there and is not collected to the BPW.
Based on this result, we modified the placement of the in-pixel circuitry  so that the electric fields  would converge toward the BPW.  We confirmed that the CCE at pixel borders is successfully improved with the updated model.
\end{Abstract}
\vfill
\begin{Presented}
International Workshop on SOI Pixel Detector\\
Sendai, Japan,  June 3--6, 2015
\end{Presented}
\vfill
\end{titlepage}
\def\thefootnote{\fnsymbol{footnote}}
\setcounter{footnote}{0}

\section{Introduction}
~~~~ Charge-coupled devices (CCDs) are widely used in X-ray astronomy, because of their fine pixel pitch ($\sim 20$~$\mu$m), low readout noise (3~e$^{-}$~rms), and high sensitivity for soft X-rays (0.3--10~keV)\cite{2001Turner}-\cite{2007Koyama}.
However, CCDs  have their own problems, such as, poor time resolution (a few seconds) and  high non-X-ray background especially above 10~keV due to high energy particles in orbit.
 To address these issues of X-ray CCDs, we have been developing active pixel sensors, referred to as ``XRPIX'',  
which  has advantages of a high-speed readout and a low background, for future X-ray astronomy satellites.

\begin{figure}[h]
\centering
\includegraphics[width=12cm,bb=0 0 1433 725]{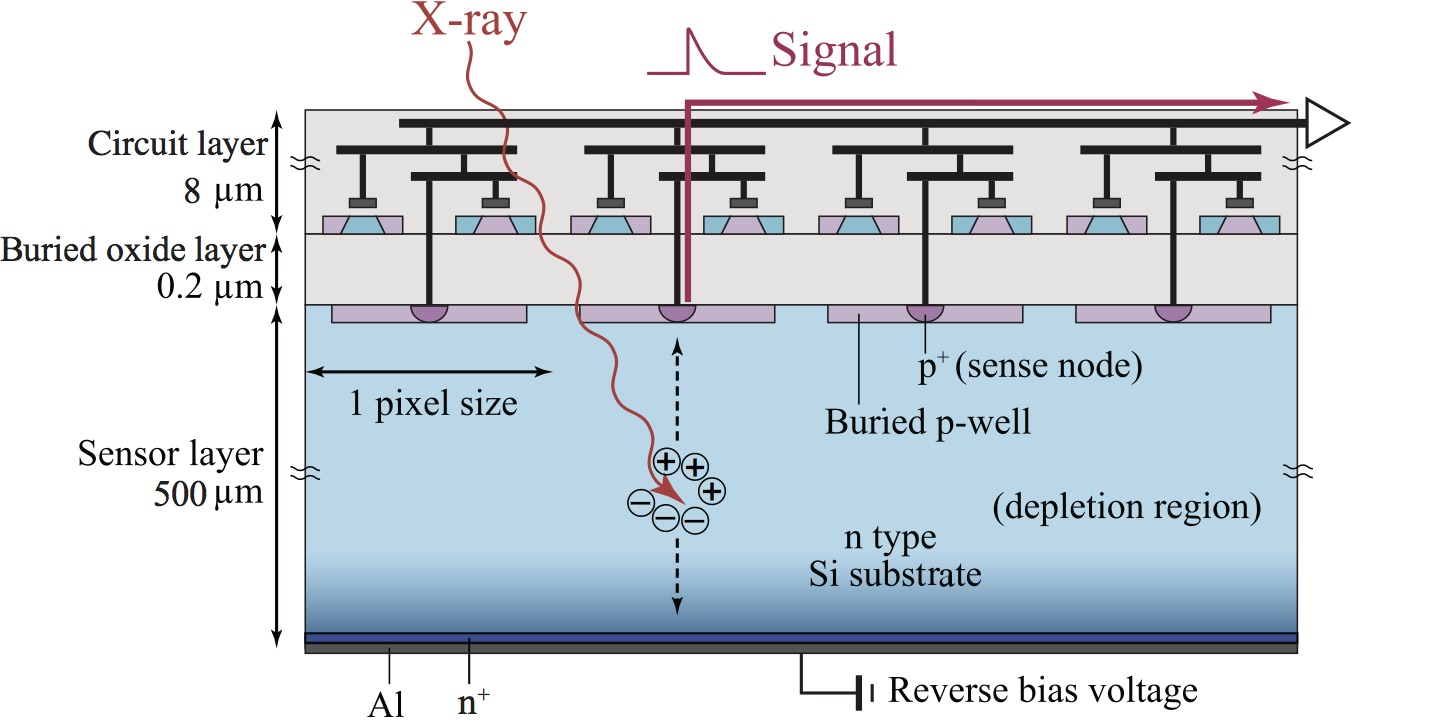}
\caption{Schematic cross-sectional view of XRPIX.}
\label{fig:CrossSection}
\end{figure}

XRPIX is fabricated, using a silicon-on-insulator (SOI) CMOS technology \cite{2011Arai},  and consists of the following three layers: 
a low-resistivity Si layer for circuits with a thickness of $\sim 8$~$\mu$m, 
a high-resistivity depleted Si layer for X-ray detection with a thickness up to 500~$\mu$m, 
and a buried oxide (BOX) layer with a thickness of $\sim 0.2$~$\mu$m for insulation between the two layers (Figure~\ref{fig:CrossSection}).
Each pixel has a sense node of p+  in the sensor layer that is connected  to the circuit through a via hole in the BOX layer. 
A buried p-well (BPW) is implemented around the sense node to suppress the back-gate effect on the circuit,  as well as to collect signal charge and then to transfer it to the sense node \cite{2011Arai}.
The pixel readout circuit has a trigger capability with the time resolution of better than 10~$\mu$sec \cite{2013Takeda}.
The in-pixel trigger circuit also  makes the event-driven readout possible.
Combining this event-driven readout and an anti-coincidence technique with surrounding scintillators,  we can  suppress the level of the non-X-ray background. 

XRPIX1 is the first prototype, the details of which are reported in Ryu et al. (2011) \cite{2011Ryu}. 
Subsequently, we developed XRPIX1b, in which  the gain was doubled, compared with XRPIX1,  by reducing the size of the BPW \cite{2013Ryu}.
  However, a significant degradation of charge-collection efficiency (CCE) was observed in XRPIX1b.
In this paper, we report  how we identified the degradation problem (Section~2),  discuss the causes of the degradation based on  our experiments (Section~3, 4), and  present the successful solution with the improved model (Section~5).
In this talk of this conference, we summarize the results from Matsumura et~al. (2014) and Matsumura et~al. (2015) \cite{2014Matsumura},\cite{2015Matsumura}.

\section{Experiment with X-rays of $^{241}$Am}
\label{sec:exp2}

\subsection{Device Description and Experimental Setup
 \label{sec:exp2_1}
 }
~~~~XRPIX1 and XRPIX1b were fabricated by using a $0.2~\mu$m fully-depleted SOI CMOS-process 
supplied by LAPIS Semiconductor Co.~Ltd.
 Both the devices have the same pixel size ($30.6~\mu{\mathrm m}\times30.6~\mu{\mathrm m}$) and format ($32\times32$ pixels),
and a high-resistivity floating zone (FZ) wafer ($\rho \geq$~1~k$\Omega$~cm).
 The differences are  the thickness of the sensor layer and the size of the  BPW.
The sensor layer is thicker in  XRPIX1b (500~$\mu$m)  than  in XRPIX1 (250~$\mu$m).
Hence, XRPIX1b has a much higher quantum efficiency in high energy X-rays than XRPIX1.
The  sizes of the BPW are $20.9~\mu{\mathrm m}\times 20.9~\mu{\mathrm m}$ in XRPIX1 and $14.0~\mu{\mathrm m}\times 14.0~\mu{\mathrm m}$ in XRPIX1b, respectively.
We expect that the node-gain  is higher in XRPIX1b  than  in XRPIX1 because  the former has a smaller parasitic capacitance.

\begin{figure}[htp]
\centering
\vspace{3mm}
\includegraphics[width=13cm,bb=0 0 1927 631]{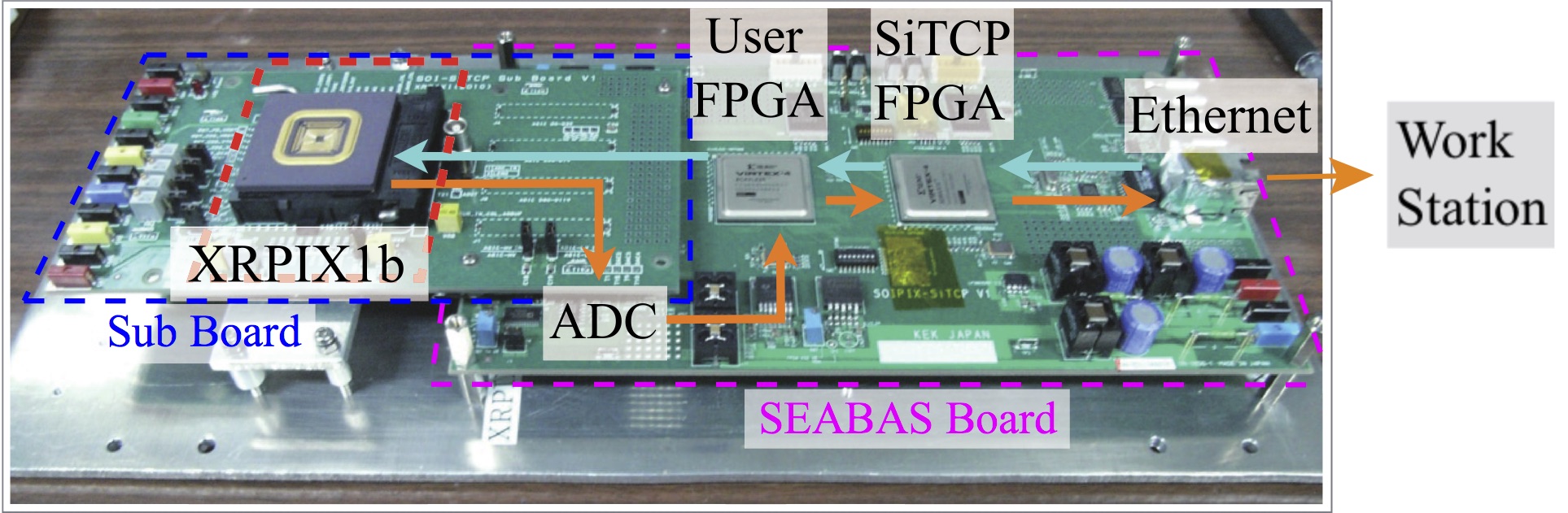}	
\caption{Photo of the experimental setup.}
\label{fig:setup}
\end{figure}

Figure~\ref{fig:setup} shows the readout system consisting of a sub board equipped with an XRPIX  and a SEABAS board \cite{2008Uchida}. 
The User FPGA on the SEABAS board generates clocks and reads out the analog signal with an ADC. 
The readout data are converted from analog to digital and then are transferred by the SiTCP FPGA to a workstation through Ethernet. 
In order to reduce the dark current, the XRPIX is cooled to $-50$~$^{\circ}$C 
in vacuum  with the pressure lower than ~$10^{-5}$ Torr.
We performed a frame-by-frame readout, where all the pixels  were sequentially read out after a 1~ms exposure.
The details of the readout sequence are presented in Ryu et al. (2011) \cite{2011Ryu}.



\subsection{Difference of Energy Spectra between XRPIX1 and 1b}
\begin{figure}[htb]
\centering
\vspace{-75mm}
\includegraphics[width=14cm,,bb=0 0 942 933]{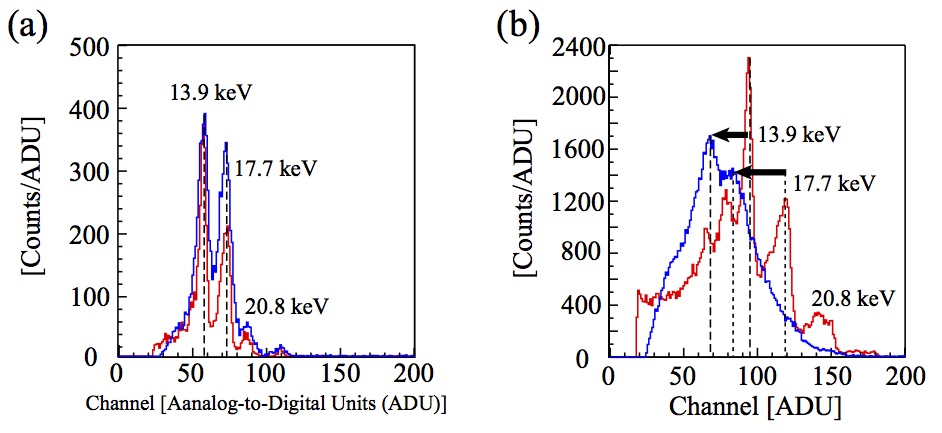}	
\caption{ Spectra of $^{241}$Am X-rays (13.9~keV, 17.7~keV and 20.8~keV) obtained with (a) XRPIX1 and (b) XRPIX1b. 
The red and blue lines denote the spectra of the single- and double-pixel events, respectively.}
\label{fig:sp_xr1_xr1b}
\end{figure}

~~~~Figure~\ref{fig:sp_xr1_xr1b} shows the spectra of $^{241}$Am X-rays at 13.9~keV, 17.7~keV, and 20.8~keV obtained  with XRPIX1 and XRPIX1b, where they were operated with  the back bias voltages of 50~V and 200~V, respectively, so that the devices achieve the full depletion.
 We define 2 event types; when the charge cloud generated with an X-ray falls into a pixel, the event is defined as a single-pixel event, and when the charge cloud spreads over two pixels, the event is a double-pixel event.

We succeeded in obtaining  roughly twice as high node-gain  in XRPIX1b  as in XRPIX1,  when peak channels  of the single-pixel events were used.
However,  new problems were immediately apparent  in the spectral shape in XRPIX1b.
Single- and double-pixel events in XRPIX1  had the same  peak channels  (Figure~\ref{fig:sp_xr1_xr1b}~left), as expected.
 On the other hand, in XRPIX1b, the peak channels  of the double-pixel events  were significantly lower than those  of the single-pixel events  (Figure~\ref{fig:sp_xr1_xr1b}~right).
We also found that the spectral shapes obtained with XRPIX1b  were distorted with large low-energy tails when compared with those obtained with XRPIX1.
These problems in XRPIX1b  were also confirmed with different back bias voltages of 30--200~V \cite{2014Matsumura}.


The distortion of the spectral shape in XRPIX1b indicates that the CCE is lower in XRPIX1b than in XRPIX1.
Nakashima et al. (2013) \cite{2013Nakashima} found that the CCE depends on the size of BPW.
Thus, problems in XRPIX1b  found here  are perhaps related to the BPW.
The shift of the peaks also suggests that the CCE is lower at the pixel boundary than that at the pixel center,  considering that the double-pixel events  occur predominantly at the pixel edge. 

\newpage

\section{Experiment with 8-keV Pencil X-ray Beams}
\subsection{Experimental Setup}
\begin{figure}[htp]
\centering
\vspace{5mm}
\includegraphics[width=13cm,bb=0 0 1552 833]{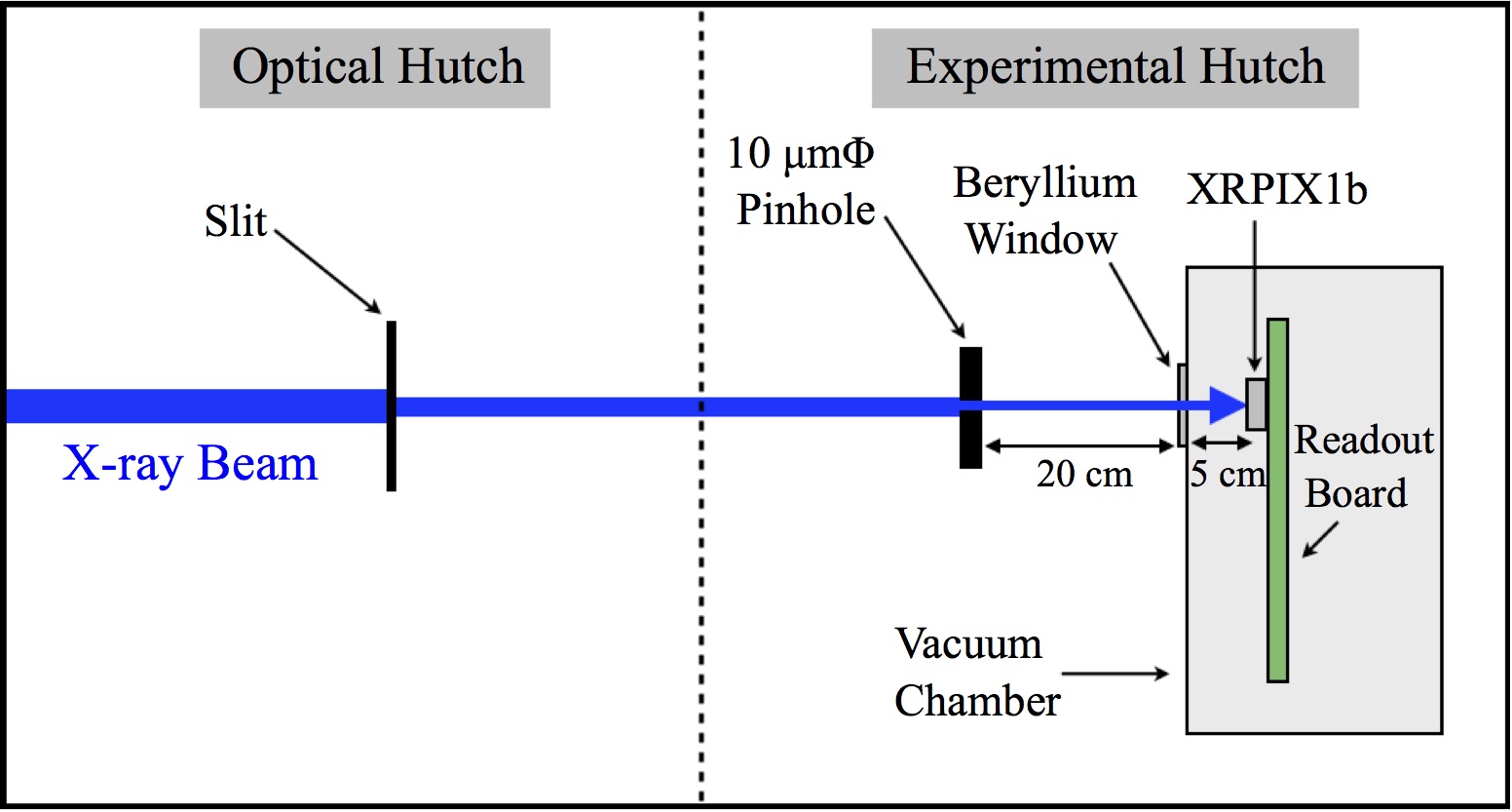}
\caption{Schematic of the experimental setup.}
\label{fig:beam_setup}
\end{figure}

~~~~In order to study the response of XRPIX1b in the sub-pixel scale, we  performed  an experiment, irradiating XRPIX1b  with finely collimated X-ray beams at 8~keV of synchrotron radiation at BL29XUL of SPring-8 \cite{2001Tamazaku}.
Figure~\ref{fig:beam_setup} shows the schematic of the experimental setup.
The beamline consists of two hutches, an optical hutch for the beam conditioning elements and an experimental hutch.
The X-ray beam is shaped with a slit in the optical hutch and a $10~\mu$m$\Phi$ pinhole placed in front of a beryllium window attached on the vacuum chamber, in which XRPIX1b and its SEABAS and sub board are installed.
The distances between the pinhole to the beryllium window and between the beryllium window and the device are 20~cm and $\leq 5$~cm, respectively.
X-rays illuminate the front-side of the device.
The XRPIX is cooled to $-50$~$^{\circ}$C in vacuum  with the pressure of lower than ~$10^{-5}$ Torr and  a bias voltage of 200~V,
which is the same as that in Section~\ref{sec:exp2}, is applied to it.
XRPIX1b on the Readout Board (Figure~\ref{fig:beam_setup}) was shifted around for both vertical and horizontal directions with a step size of 6um in each step so that the X-ray beam made a grid scan over a pixel in XRPIX1b, each of which has the size of $30.6~\mu{\mathrm m}\times30.6~\mu{\mathrm m}$ (Section~\ref{sec:exp2_1}).
At each grid point, the events were recorded with a frame-by-frame readout.

\subsection{Result of the SPring-8 Beam Experiment}
\begin{figure}[h]
\centering
\vspace{10mm}
\includegraphics[width=15cm,bb=0 0 3893 1214]{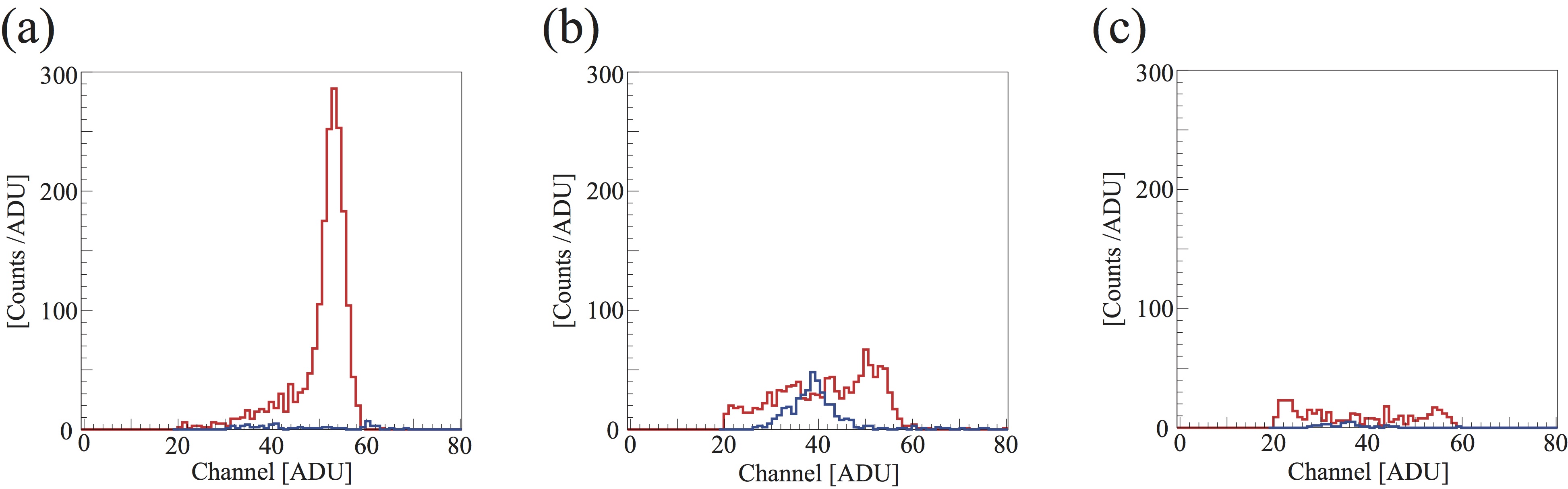}
\caption{The spectra of XRPIX1b obtained by irradiating with 8.0~keV X-ray beams at (a) pixel center, (b) boundary, and (c) corner, respectively.
The red and blue lines denote spectra of single- and double-pixel events, respectively.}
\label{fig:CCE}
\end{figure}

~~~~ Figure~\ref{fig:CCE} shows the spectra of XRPIX1b,  where it was irradiated with 8.0~keV X-ray beams at pixel center, boundary, and corner.
Red and blue lines denote spectra of single- and double-pixel events, respectively.
In the spectra at the pixel center, we see a single line as expected. 
However, in the spectra at the pixel border, the peak energies of double-pixel events are significantly lower than those of single-pixel events.  Worse, the spectra at the pixel corner  show no line.
 This results well demonstrates that XRPIX1b has dead regions at the pixel borders and corners because the CCE degrades.

\subsection{Cause of Degradation of the CCE}
~~~~The experimental result  presented in Section~\ref{sec:exp2} and Nakashima et al. (2013) \cite{2013Nakashima} suggest that the CCE is reduced as the BPW size is reduced with respect to the pixel size.
In addition, our experiment shows that the CCE depends on the position  within a pixel, namely the CCE is significantly lower at the pixel boundary than  at the pixel center. 
The pixel boundary, which is not covered by the BPW, has numerous crystal defects at the interface of the sensor and BOX layers. 
Therefore, a part of the signal charge which moves to the pixel boundary is possibly lost.

\newpage

\section{Simulation of Electric Fields and Potentials}
\label{sec:sim1and1b}
\subsection{Simulation  for XRPIX1}
\begin{figure}[htbp]
\centering
\includegraphics[width=14.5cm,,bb=0 0 3147 1139]{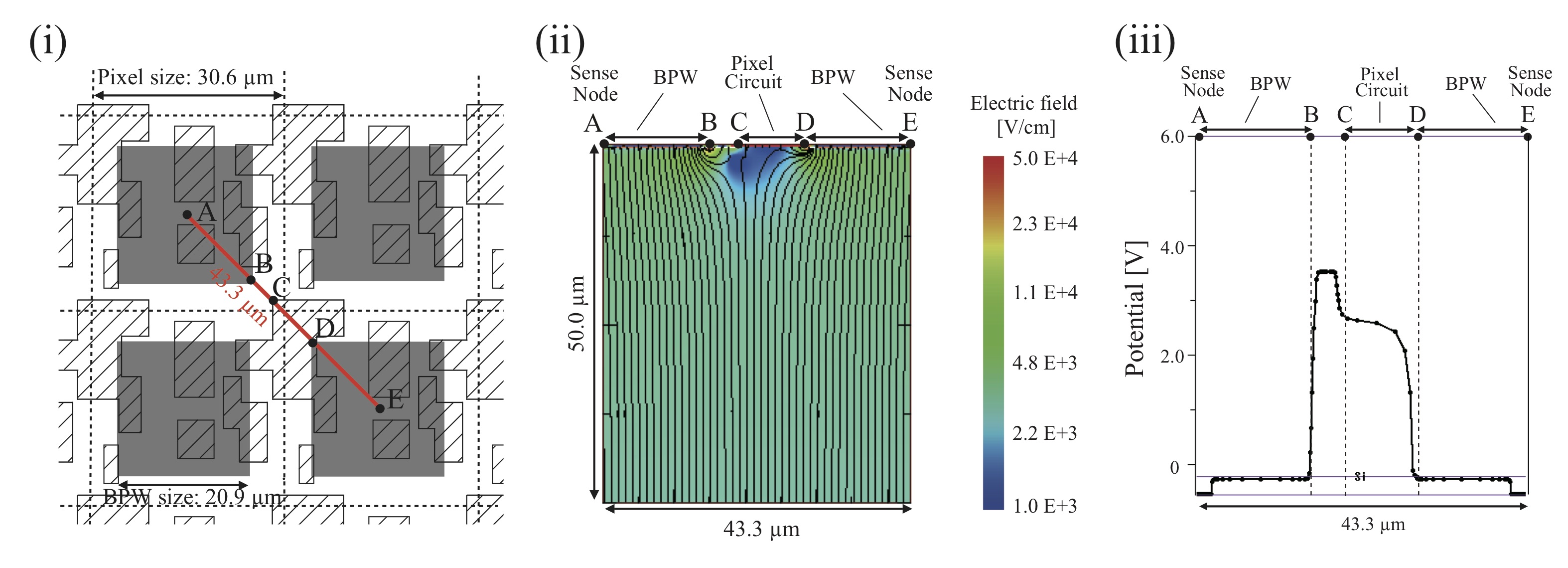}	
\caption{Configuration and results of the simulation for XRPIX1. (i) The pixel layout of XRPIX1. The gray and hatched areas indicate the BPWs and circuit locations, respectively.
(ii) Simulation of the electric fields in the sensor layer along the red line shown in (i). 
(iii) Simulation of the electric potentials in the sensor layer close to the interface with the BOX layer. 
The positions of A to  E in (ii) and (iii) correspond to those in (i).
}
\label{fig:Simu_1_fc}
\end{figure}

~~~~In order to confirm route which signal charge pass through, we ran a 2D simulation of the electric fields and potentials.
We used the semiconductor device simulator HyDeLEOS, which is a part of the TCAD system HyENEXSS \cite{HyENEXSS}.
Region of this simulation is cross-section that is vertical to straight line A-E in Figure~\ref{fig:Simu_1_fc}~(i), which connects two sense nodes.
We fixed the electric potentials of the sense nodes, the BPWs, and pixel circuits at 0~V, and the back bias at 50~V.
A fixed positive charge is generated in the BOX during the wafer process.
We assumed the typical fixed positive charge to be $2.0 \times 10^{11}~\rm{cm^{-2}}$, based on \cite{1998Afanas},\cite{PixelDetector}.
As the initial conditions, the charge was given to be uniformly distributed in the region between 1 nm and 3 nm above the sensor-BOX interface.

Figure~\ref{fig:Simu_1_fc}~(ii) and (iii) show the resultant electric fields and potentials obtained with the simulation, respectively. 
The potentials are close to the sensor-BOX interface.
The former is found to penetrate into the region of the circuitry, whereas those under A-B and D-E converge into the BPWs. 
The signal charge under A-B or D-E moves to the BPWs and is detected by the sense node without any significant loss.
The charge under C-D drifts to the region under the pixel circuitry outside the BPWs,
 and is nevertheless detected, because the electric potential under C-D inclines toward the BPWs.

\subsection{Simulation for XRPIX1b}
\begin{figure}[htbp]
\centering
\includegraphics[width=14.5cm,bb=0 0 3147 1154]{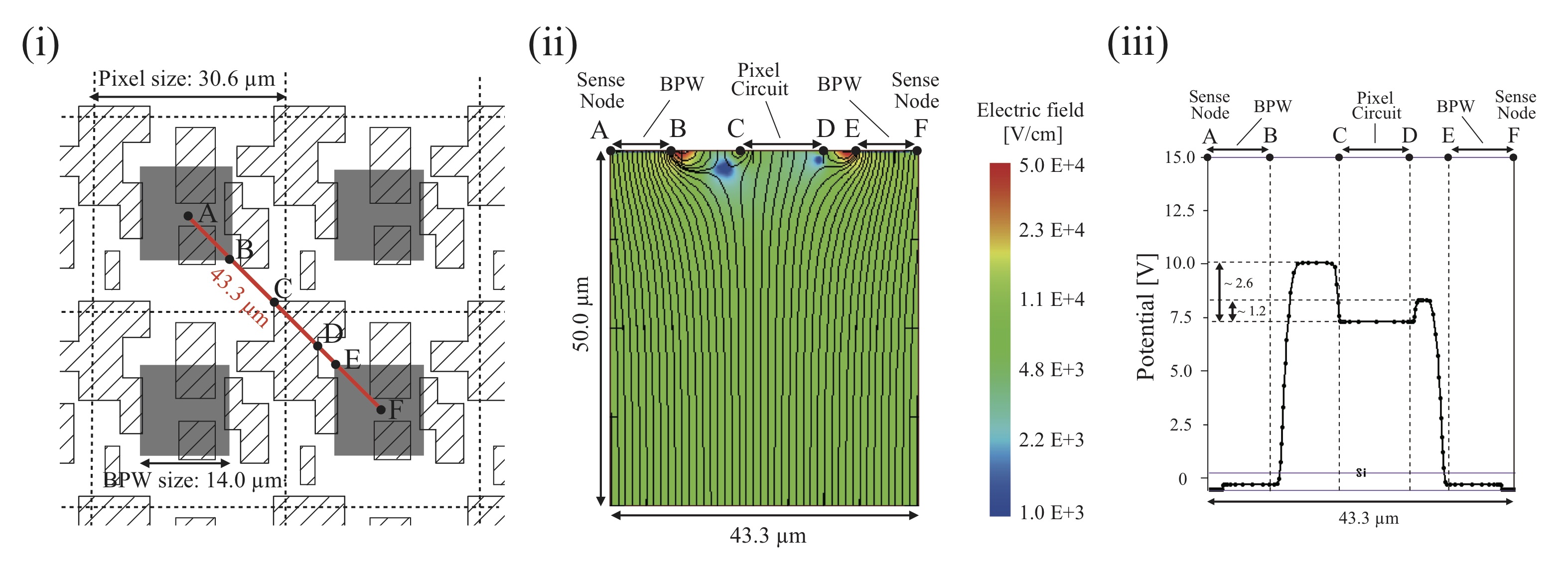}
\caption{Configuration and results of the simulation for XRPIX1b.  See Figure~\ref{fig:Simu_1_fc}  for notations.} 
\label{fig:Simu_1b_fc}
\end{figure}

~~~~We also ran a simulation  for XRPIX1b, using  almost the same parameters  as for XRPIX1.
 The differences in the parameters for the simulations between XRPIX1 and 1b are the sizes of the BPWs, thickness of the sensor layer, and back bias voltage (200~V for 1b)  (Figure~\ref{fig:Simu_1b_fc}~(i)). 
 We found the stark contrast in the electric-field structure under C-D with that for XRPIX1 (Figure~\ref{fig:Simu_1_fc}~(ii, iii) and Figure~\ref{fig:Simu_1b_fc}~(ii, iii)).  That for XRPIX1b shows the clear local minimum, which does not exist in that for XRPIX1.
The potential barriers are 2.6~V and 1.2~V on the C and D sides, respectively. 
The thermal energy of a hole is $\sim 0.02$~eV at $-50$~$^{\circ}$C.
Therefore, a hole cannot easily escape from the local minimum, and then a significant part of the signal charge is subsequently lost.  It is probably trapped by the surface defects at the sensor-BOX interface. 
Consequently, this must be the  cause of the degradation of the CCE in XRPIX1b, that is,  the local minimum in the electric potential created  with the pixel circuitry, which is located far from the BPWs.

\section{Improvement of CCE}
~~~~In the study with XRPIX1 and 1b, we found that the location of the pixel circuitry in a pixel has a significant effect on charge collection.
It is necessary to solve the problem without increasing the size of BPWs in order to keep the parasitic capacitance of  pixels low.
We realized that it would be possible to control the electric fields by adjusting the location of the pixel circuitry.
With an appropriate layout of the circuitry,  the electric fields would converge into BPWs and X-ray events would be detected without any significant loss in the signal charge.
We have then developed the device, named XRPIX2b, of which the pixel circuitry is located near the BPWs in order to improve the CCE.

\subsection{Simulation of Electric Field in XRPIX2b}
~~~~XRPIX2b is the fourth device that we have developed. 
XRPIX2b has almost the same pixel size of 30.0~$\mu$m~$\times$ 30.0~$\mu$m 
as XRPIX1 and 1b,  whereas the BPW size is 12.0~$\mu$m~$\times$~12.0~$\mu$m, which is somewhat smaller than 20.9~$\mu$m~$\times$~20.9~$\mu$m of XRPIX1 and 14.0~$\mu$m~$\times$~14.0~$\mu$m of XRPIX1b.
The sensor layer of XRPIX2b has the same thickness as that of XRPIX1b (500~$\mu$m) \cite{D_Takeda}.

\begin{figure}[htbp]
\centering
\includegraphics[width=15cm,bb=0 0 3197 1107]{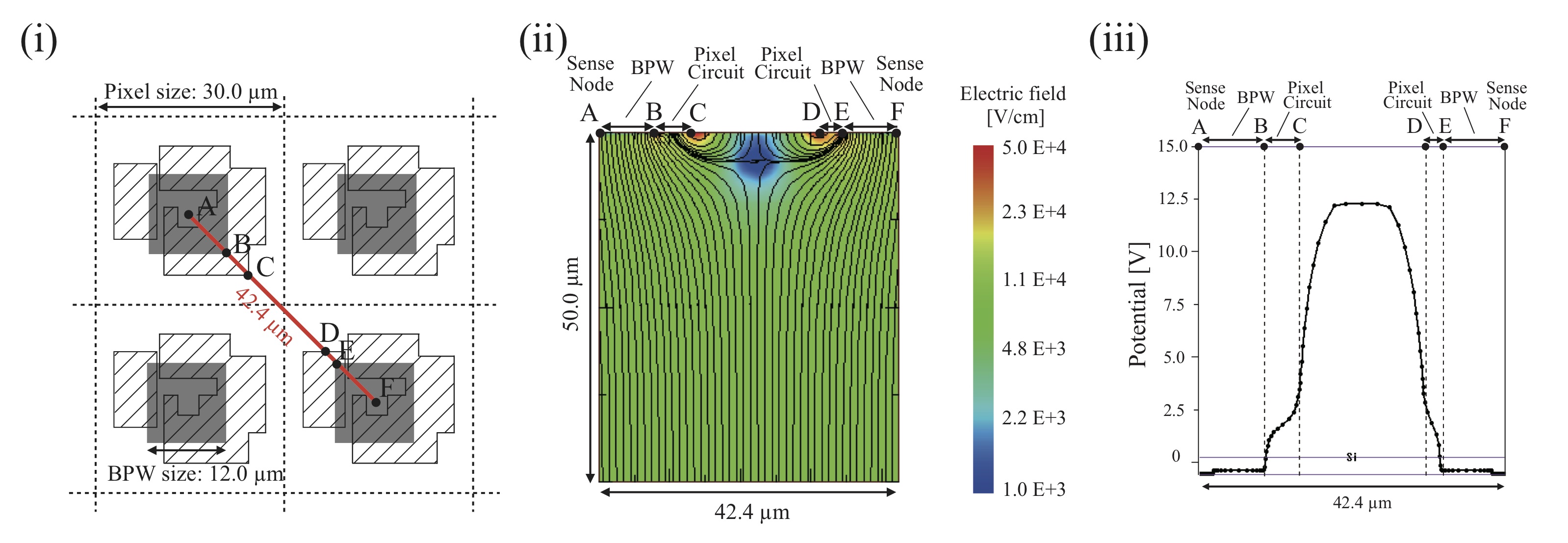}
\caption{Configuration and results of the simulation for XRPIX2b.  See Figure~\ref{fig:Simu_1_fc}  for notations.}
\label{fig:Simu_2b_fc}
\end{figure}

In XRPIX2b, we placed the in-pixel circuitry near the BPWs, as shown in Figure~\ref{fig:Simu_2b_fc}~(i).
In contrast to XRPIX1b, the pixel circuitry is confined to the central part of the pixel.
Figure~\ref{fig:Simu_2b_fc}~(ii) gives the  simulated results of the electric fields, where we assumed the electric potentials for the pixel circuits, the BPWs, and the sense nodes to be 0 V, as in the case for the XRPIX1b simulation (Section~\ref{sec:sim1and1b}).
The electric field lines in Figure 8(ii) indicate that signal charge generated in the sensor layer is swept to the regions A-C or D-F.
The charge carried to A-B or E-F reaches the BPWs and is detected without loss.
On the other hand, the charge moved to the regions B-C or D-E, which are outside the BPWs, is expected to drift without contacting the BOX-sensor interface.
Figure~\ref{fig:Simu_2b_fc}~(iii) shows the simulated electric potentials at the interface.
The local minimum observed in the XRPIX1b case has disappeared. 
Instead, sharp  gradients of the electric potential toward the BPWs are formed.
These imply that the signal charge carried to B-C or D-E (or even C-D) will drift to the BPWs quickly and with minimum loss.
Thus, we don't expect  much degradation of the CCE in XRPIX2b.

\subsection{X-ray Irradiation Experiment of XRPIX2b}
\begin{figure}[htb]
\centering
\includegraphics[width=8cm,bb=0 0 443 431]{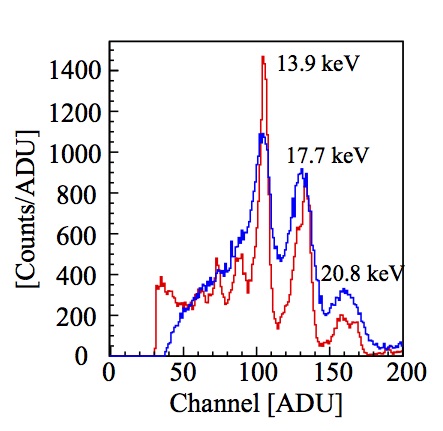}	
\caption{Spectra of $^{241}$Am X-rays obtained with XRPIX2b. 
Note that the red and blue lines denote the spectra of the single- and double-pixel events, respectively.
Besides the three lines from $^{241}$Am (13.9~keV, 17.7~keV, and 20.8~keV), 
three lines of Cu~K$\alpha$ at 8.0~keV, Au~L$\alpha$ at 9.7~keV, and Au~L$\beta$ at 11.4~keV are also seen in the spectra.
They are fluorescence X-rays from the cold plate or a gold palladium alloy window of the $^{241}$Am radioisotope.
}
\label{fig:sp_xr2b}
\end{figure}

~~~~Now we examine the performance of XRPIX2b with real X-rays; we made the same experiment with 241Am as those for its predecessors (Section~\ref{sec:exp2}).
Figure~\ref{fig:sp_xr2b} shows the spectra of $^{241}$Am X-rays obtained with XRPIX2b, in which the red and blue lines are single- and double-pixel events, respectively.
The frame-by-frame mode was used for the readout.
The devices were cooled down to $-50$~$^{\circ}$C and the back bias voltage of 200~V was applied. 

We confirmed that the peak energies of double-pixel events are significantly lower than those of single-pixel events in the spectra of XRPIX1b (Figure~\ref{fig:sp_xr1_xr1b}~(b)).
This is consistent with the result that the CCE degrades at the pixel borders.
We found that  in XRPIX2b, the peak positions in the double-pixel events spectrum coincide with those in the single-pixel events spectrum, being very different from that in XRPIX1b. This indicates no degradation of the CCE at the pixel borders in XRPIX2b. 
This result  implies that we successfully control the electric fields and potentials in XRPIX2b by placing in-pixel circuitry so that  signal charges generated even at a pixel border  are collected without loss. 

The guideline of the layout we  have learned in this study is to confine the in-pixel circuitry located close to the BPW in order not to lose the signal charge. 
This means that a larger circuit size requires a larger area of the BPW, which leads to an increase in the parasitic capacitance of the node and reduction of the gain.
One of possible solutions is an introduction of a high-gain amp at the sense node.
We give a relevant report  about this issue in Takeda et al. (2015) \cite{2015Takeda}.

\section{Summary}
~~~~We found that the detection efficiency and the CCE of XRPIX1b degrade at the position 
under the pixel circuitry with the experiment of the X-ray beams, whose diameters are $10~\mu$m$\Phi$. 
A 2D simulation shows that the pixel circuitry outside the BPW makes local minimums in the electric potentials in the sensor layer close to the interface with the BOX layer. 
These local minimums should be responsible for  the degradation of the CCE,  because the signal charge  that are created near the local minimum is difficult to escape from there and thus a significant part of charge is lost. 
It is possible to control the electric fields and potentials by placing the circuitry appropriately. 
We rearranged the in-pixel circuitry in XRPIX2b so that the electric fields converge toward the BPWs and successfully improved the CCE at pixel borders.

\newpage

\Acknowledgements
We would like to acknowledge the valuable advice and great work by the personnel of LAPIS Semiconductor Co., Ltd.
Use of BL29XUL at SPring-8 was supported by RIKEN.  
This work is supported by JSPS KAKENHI grant number 15J01842 (H.M.).
This work is also supported by JSPS Scientific Research grant numbers 25109002 (Y.A.), 23340047 and 25109004 (T.G.T.), 25109004 and 25870347 (T.T.), and 15K17648 (A.T.).
We express our thanks to the development team of HyENEXSS.

\end{document}